\documentclass[sigconf]{acmart}

\usepackage{multirow} 
\usepackage{array}
\usepackage{makecell}
\usepackage{url}
\usepackage{enumitem}

\newcommand{\sota}{SOTA}

\newcommand{\bfred}[1]{{\color{red}{\textbf{#1}}}}

\newcommand{\bfblue}[1]{{\color{blue}{\textbf{#1}}}}

\usepackage{etoolbox}
\makeatletter
\patchcmd{\authornote}{\g@addto@macro\addresses{\@authornotemark}}{}{}{}
\makeatother

\AtBeginDocument{%
  \providecommand\BibTeX{{%
    \normalfont B\kern-0.5em{\scshape i\kern-0.25em b}\kern-0.8em\TeX}}}

\setcopyright{acmcopyright}
\copyrightyear{2018}
\acmYear{2018}
\acmDOI{}

\acmConference[Conference acronym 'XX]{Make sure to enter the correct
  conference title from your rights confirmation email}{May 22, 2023}{Woodstock, NY}
\acmPrice{15.00}
\acmISBN{978-1-4503-XXXX-X/18/06}

\begin{document}


\title{Automatic Code Summarization via ChatGPT: How Far Are We?}


\author{Weisong Sun$^{1,2}$, Chunrong Fang$^{1}$*, Yudu You$^1$, Yun Miao$^1$, Yi Liu$^2$, Yuekang Li$^3$, Gelei Deng$^2$, Shenghan Huang$^1$, Yuchen Chen$^1$, Quanjun Zhang$^1$, Hanwei Qian$^1$, Yang Liu$^2$, Zhenyu Chen$^1$}
\authornote{Corresponding author.}
\affiliation{%
  \institution{$^1$State Key Laboratory for Novel Software Technology, Nanjing University, Nanjing, China \\ $^2$Nanyang Technological University, Singapore \\ $^3$ University of New South Wales, Australia}
  \country{}
}
\email{*fangchunrong@nju.edu.cn}

\renewcommand{\shortauthors}{Weisong Sun and Chunrong Fang, et al.}

\begin{abstract}
To support software developers in understanding and maintaining programs, various automatic code summarization techniques have been proposed to generate a concise natural language comment for a given code snippet. Recently, the emergence of large language models (LLMs) has led to a great boost in the performance of natural language processing tasks. Among them, ChatGPT is the most popular one which has attracted wide attention from the software engineering community. However, it still remains unclear how ChatGPT performs in (automatic) code summarization. Therefore, in this paper, we focus on evaluating ChatGPT on a widely-used Python dataset called CSN-Python and comparing it with several state-of-the-art ({\sota}) code summarization models. Specifically, we first explore an appropriate prompt to guide ChatGPT to generate in-distribution comments. Then, we use such a prompt to ask ChatGPT to generate comments for all code snippets in the CSN-Python test set. We adopt three widely-used metrics (including BLEU, METEOR, and ROUGE-L) to measure the quality of the comments generated by ChatGPT and {\sota} models (including NCS, CodeBERT, and CodeT5). The experimental results show that in terms of BLEU and ROUGE-L, ChatGPT’s code summarization performance is significantly worse than all three {\sota} models. We also present some cases and discuss the advantages and disadvantages of ChatGPT in code summarization. Based on the findings, we outline several open challenges and opportunities in ChatGPT-based code summarization.
\end{abstract}



\maketitle

\section{Introduction}
\label{sec:introduction}
Code comments play a critical role in facilitating program comprehension~\cite{1981-Comments-on-Program-Comprehension, 1988-Program-Readability} and software maintenance~\cite{1993-Maintenance-Productivity, 2005-Documentation-Essential-Software-Maintenance}. It is a good programming practice to write high-quality code comments~\cite{2005-Documentation-Essential-Software-Maintenance, 2020-CPC, 2022-EACS}.  
However, writing comments is a labor-intensive and time-consuming task. Consequently, high-quality comments are often absent, unmatched, and outdated during software evolution. Lack of high-quality code comments is a common problem in the software industry~\cite{2018-TL-CodeSum, 2022-Practitioners-Expectations-on-Comment-Generation}. 
Automatic code summarization (code summarization, for short) is a hot research topic~\cite{2021-Neural-Code-Summarization-How-Far, 2021-Why-My-Code-Summarization-Not-Work, 2022-Evaluation-Neural-Code-Summarization}, which aims to design advanced techniques to support the automatic generation of code summaries (i.e., comments). For a code snippet (e.g., a Java method or a Python function) given by the developer, code summarization techniques can automatically generate natural language summaries for it.

In recent years, a large number of deep learning (DL)-based code summarization techniques have been proposed one after another, such as DeepCom~\cite{2018-DeepCom}, NCS~\cite{2020-Transformer-based-Approach-for-Code-Summarization}, and SIT~\cite{2021-SiT}. They leverage powerful generative models trained on a large-scale code-comment corpus to translate code snippets in programming languages into summaries in natural language~\cite{2020-Code-to-Comment-Translation, 2022-Evaluation-Neural-Code-Summarization, 2019-Ast-attendgru, 2018-Code2seq, 2018-DeepCom, 2016-CODE-NN}. Leveraging powerful deep learning, these techniques have shown to outperform traditional information retrieval (IR)-based techniques~\cite{2010-Program-Comprehension-with-Code-Summarization, 2010-Towards-Generating-Summary-Java-Methods, 2010-Automated-Text-Summarization-Summarizing-Code, 2013-Automatic-Generation-Summaries-for-Java-Classes, 2014-Code-Summarization-of-Method-Context} in terms of the naturalness and informativeness of the generated summaries~\cite{2018-DeepCom, 2020-R2Com, 2021-EditSum}. After that, with the success of the pre-training and fine-tuning paradigm in the field of NLP (e.g., BERT and T5), many works in software engineering (SE) introduce this paradigm to further boost code-related tasks, including code summarization (e.g, CodeBERT~\cite{2020-CodeBERT} and CodeT5~\cite{2021-CodeT5}). In practice, these works first pre-train a large-scale language model (i.e., LLM) with general language modeling tasks, such as masked language modeling (MLM) and unidirectional language modeling (ULM). Then, they fine-tune the pre-trained models on downstream tasks, such as code clone detection, code search, and code summarization. 

Most recently, an intelligent human-machine dialogue LLM called ChatGPT has attracted great attention. ChatGPT is created by fine-tuning a GPT-3.5 series model via reinforcement learning from human feedback (RLHF)~\cite{2017-Reinforcement-Learning-Human-Preferences}. Compared with previous DL approaches, ChatGPT accepts longer input and enables users to have a conversation with the language model while the previous chat history is taken into account when generating answers. With the emergence of ChatGPT, there is growing interest in leveraging this model for various NLP tasks~\cite{2023-ChatGPT-NLP-task-solver, 2023-ChatGPT-Good-Translator, 2023-ChatGPT-Multitask-Multilingual-Multimodal-Evaluation, 2023-ChatGPT-for-Aspect-based-Text-Summarization}. In addition to natural language, ChatGPT can also deal with source code, which arouses growing interest in applying ChatGPT to SE tasks~\cite{2023-Prompt-Design-of-ChatGPT-for-Program-Repair, 2023-Bug-Fixing-of-ChatGPT, 2023-Investigating-Code-Generation-of-ChatGPT, 2023-Code-Generation-via-ChatGPT}. However, the exploration of ChatGPT in code summarization is still lacking.

\begin{figure}[t]
  \centering
  \includegraphics[width=\linewidth]{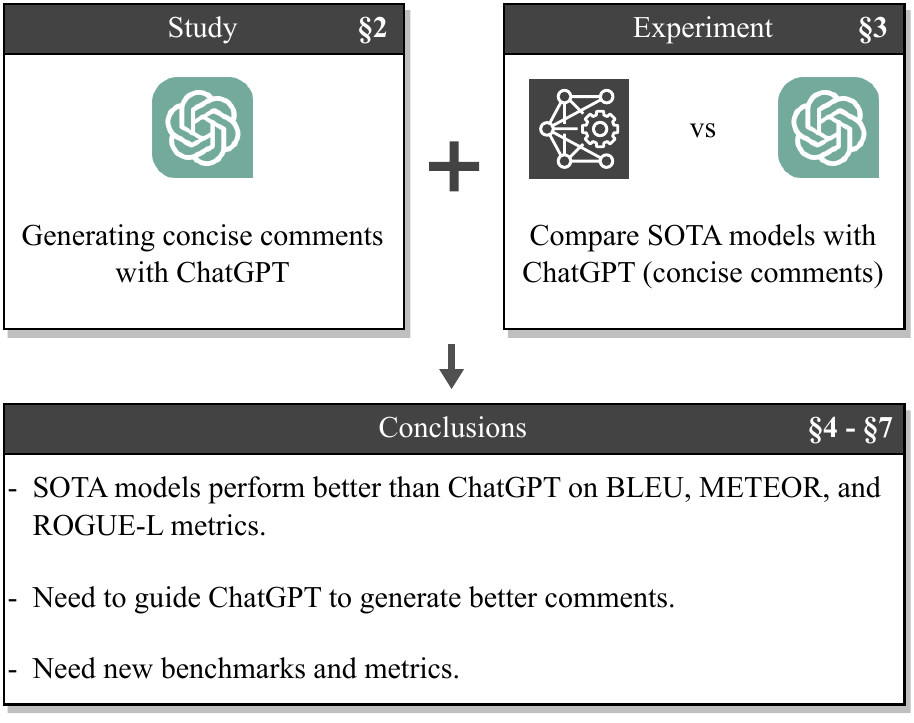}
  \caption{Overview of this paper}
  \label{fig:overview}
\end{figure}

In this paper, we perform a preliminary evaluation of ChatGPT's zero-short code summarization performance. 
The overview of this study is depicted in Figure~\ref{fig:overview}.
The overall idea is to explore the potential of ChatGPT in code summarization and compare the explored strategy of using ChatGPT with state-of-the-art (\sota{}) techniques to draw conclusions.
Specifically, we first design several heuristic questions to collect the feedback of ChatGPT. 
Based on the feedback, we find an appropriate prompt to guide ChatGPT to generate in-distribution comments. 
Then, we use such a prompt to ask ChatGPT to generate comments for all 14,918 code snippets in the CSN-Python test set. We leverage three widely-used metrics (i.e., BLEU, METEOR, ROUGE-L) to measure the quality of the comments generated by ChatGPT compared to ground-truth comments. Moreover, to provide a deeper understanding of ChatGPT's code summarization performance, we provide a comparison with results obtained by three \sota{} code summarization models (including NCS~\cite{2020-Transformer-based-Approach-for-Code-Summarization}, CodeBERT~\cite{2020-CodeBERT}, and CodeT5~\cite{2021-CodeT5}). The experimental results show that overall ChatGPT is inferior to the three \sota{} code summarization models in terms of BLEU, METEOR, and ROUGE-L. We also show and analyze some successful, moderate, and weak cases of ChatGPT. Based on the findings, we outline several challenges and opportunities in ChatGPT-based code summarization that remain to be addressed in future work, including how to improve ChatGPT on existing benchmark datasets and metrics, how to crop the rich but lengthy comments generated by ChatGPT, how to create a higher-quality benchmark dataset by combing existing datasets with ChatGPT-generated comments, and how to design suitable metrics to evaluate ChatGPT-based code summarization models.

In summary, we make the following contributions.
\begin{itemize}
    \item To the best of our knowledge, we are the first to investigate and analyze ChatGPT's code summarization performance.

    \item We find an appropriate prompt to guide ChatGPT to generate in-distribution comments through carefully crafted heuristic questions and feedback.
    
    \item We evaluate ChatGPT on a widely-used code summarization dataset and compare it with three {\sota} code summarization models.

    \item Based on the findings, we outline several challenges and opportunities for ChatGPT-based code summarization.

\end{itemize}

\section{ChatGPT for Automatic Code Summarization}
\label{sec:ChatGPT_for_code_summarization}
In this section, we present five carefully crafted heuristic questions for exploring the performance of ChatGPT~\footnote{Note that in this paper, we interact with ChatGPT by calling the officially provided API.} in code summarization. The five heuristic questions are as follows:
\begin{itemize}[left=1em]
    \item[\textbf{Q1.}] Can ChatGPT perform code summarization tasks? 
    \item[\textbf{Q2.}] What does the comment generated by ChatGPT look like?
    \item[\textbf{Q3.}] How to use ChatGPT to generate concise comments?
    \item[\textbf{Q4.}] What type of prompt does ChatGPT suggest for generating short comments?
    \item[\textbf{Q5.}] Which one performs better, the ChatGPT-suggested prompts in Q4 or the prompt proposed in Q3?
\end{itemize}
Based on observations of answers (feedback) from ChatGPT, we provide an appropriate prompt that guides ChatGPT to generate in-distribution comments. Detailed questions, answers, and observations are described in the following sections.

\subsection{Q1: Can ChatGPT perform code summarization tasks?}

\begin{figure}[htbp]
  \centering
  \includegraphics[width=\linewidth]{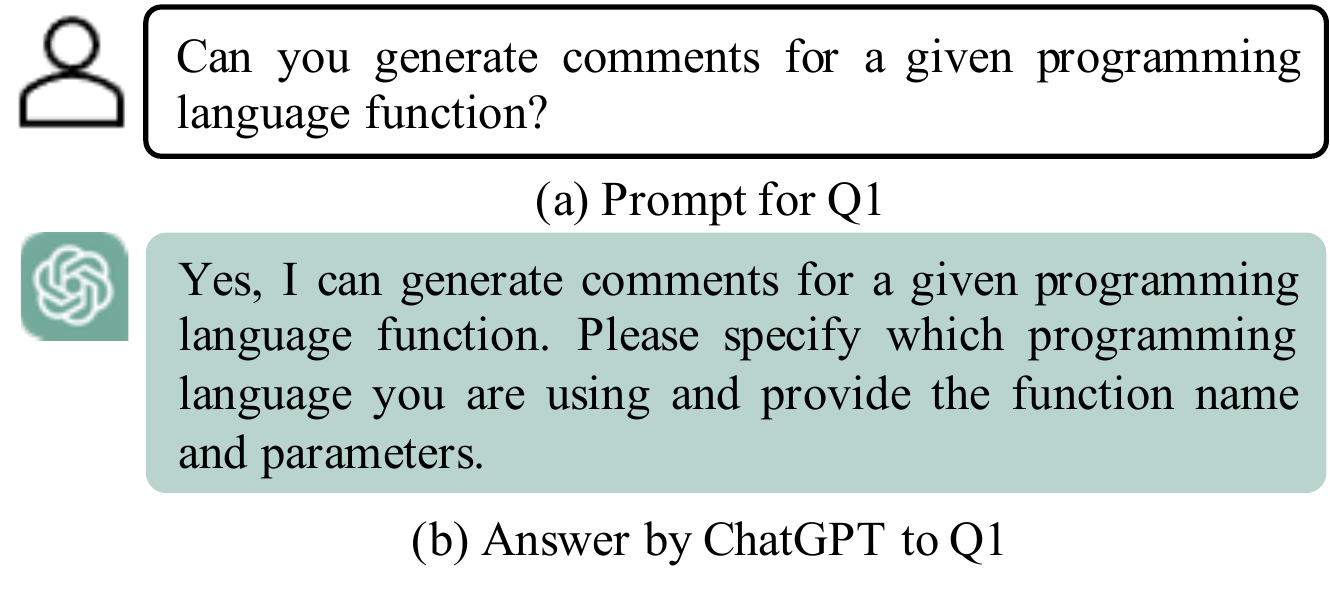}
  \caption{Prompt for Q1 and answer by ChatGPT to Q1}
  \label{fig:prompt_answer_Q1}
\end{figure}

In this question, we want to figure out whether ChatGPT can be used for code summarization. Figure~\ref{fig:prompt_answer_Q1}(a) and (b) show the prompt we used to query ChatGPT and the answer of ChatGPT, respectively.

From Figure~\ref{fig:prompt_answer_Q1}(b), it is observed that ChatGPT clearly answered that it can generate comments for a given programming language function. Therefore, we can conclude that although ChatGPT is not primarily intended for automated code summarization, it can still be applied to it. 

\begin{figure}[htbp]
  \centering
  \includegraphics[width=\linewidth]{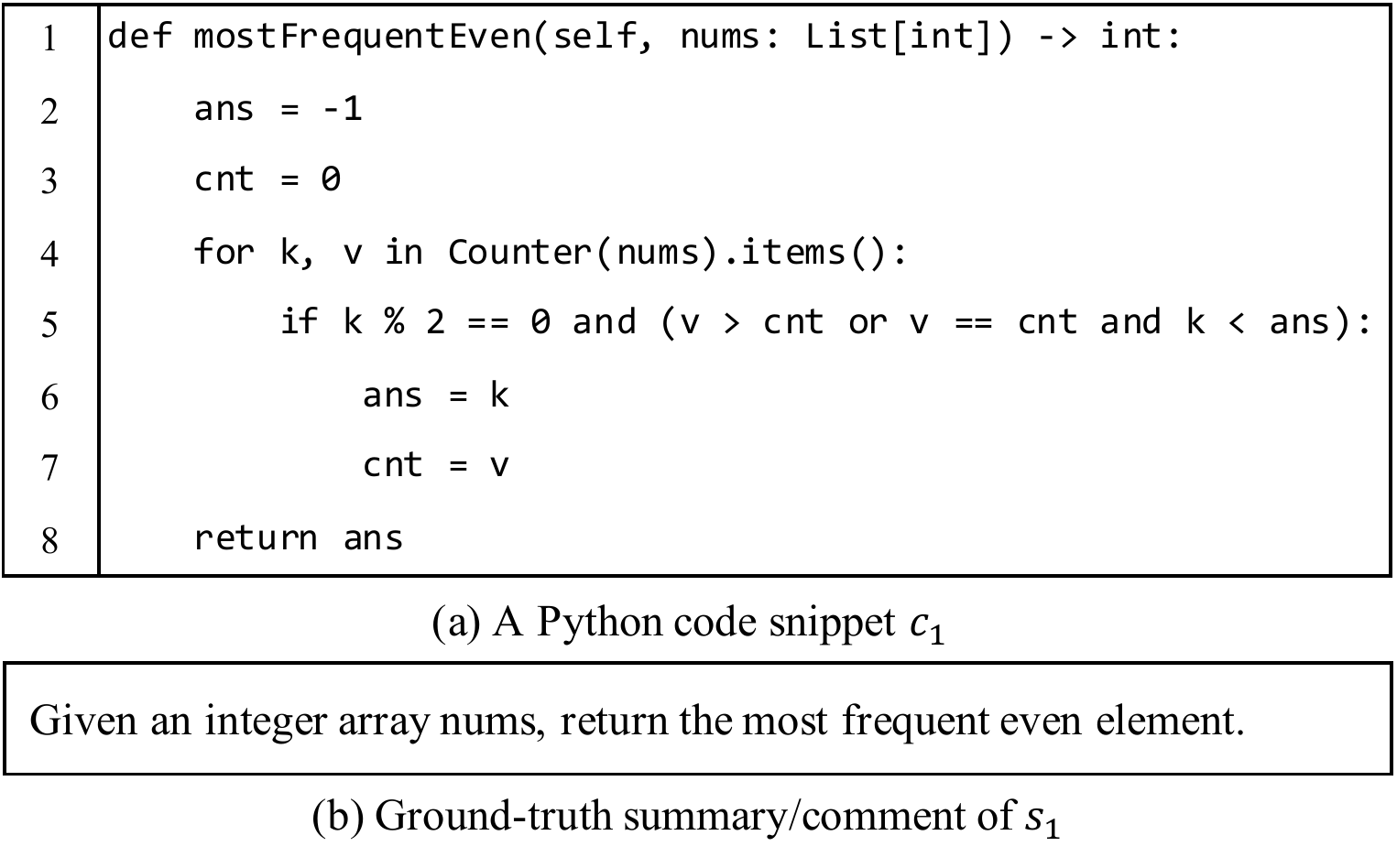}
  \caption{An example of code snippet and ground-truth summary/comment}
  \label{fig:example_of_code_ground-truth_comment}
\end{figure}

\begin{figure*}[!t]
  \centering
  \includegraphics[width=\linewidth]{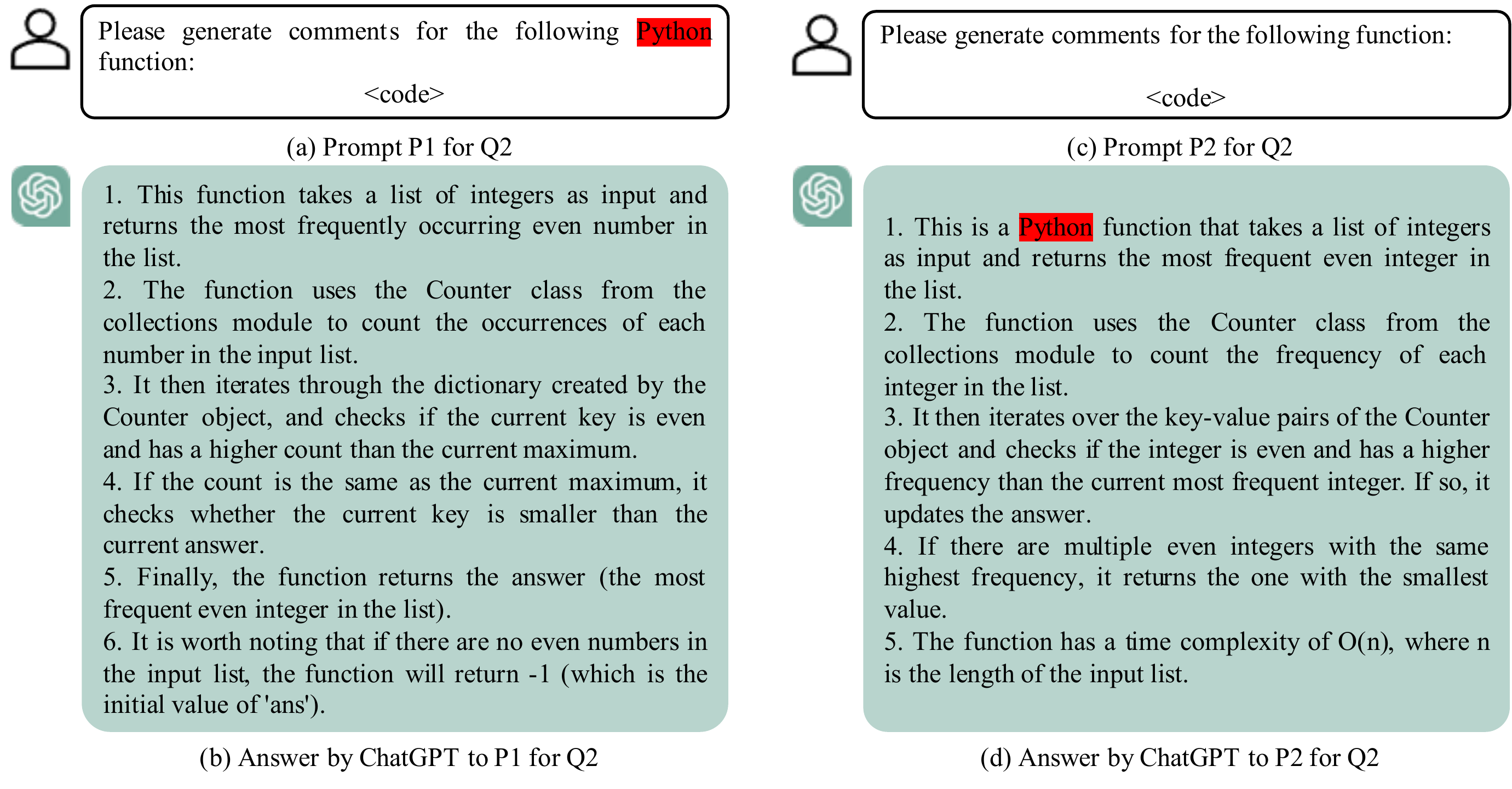}
  \caption{Two prompts for Q2 and the corresponding answers by ChatGPT to Q2. The ``<code>'' in the figure represents a placeholder, and the code snippet we filled in during the experiment is $c_1$ shown in Figure~\ref{fig:comment_number_by_sentence_number}(a).}
  \label{fig:prompt_answer_Q2}
\end{figure*}

\subsection{Q2: What does the comment generated by ChatGPT look like?}

In this question, we want to take a look at the comments ChatGPT generates for a specific code snippet. We take the code snippet $c_1$ in Figure~\ref{fig:example_of_code_ground-truth_comment}(a) as an example and ask ChatGPT to generate comments for it. Figure~\ref{fig:prompt_answer_Q2} show the two prompts we used to ask ChatGPT and the corresponding answers by ChatGPT. To avoid unnecessary redundancy, in Figure~\ref{fig:prompt_answer_Q2}, we use the placeholder ``<code>'' to refer to the code snippet $c_1$. From Figure~\ref{fig:prompt_answer_Q2}(b), it is observed that, compared with the ground-truth comment $s_1$ shown in Figure~\ref{fig:example_of_code_ground-truth_comment}, the comments generated by ChatGPT for $c_1$ are very detailed. 

Intuitively, different programming languages (e.g., Java and Python) have obvious differences in syntax, style, and coding standards. Therefore, by the way, we verified whether it is possible to not specify a programming language under the premise of providing a code snippet. Figure~\ref{fig:prompt_answer_Q2}(c) shows an example. Compared with the prompt P1 in Figure~\ref{fig:prompt_answer_Q2}(a), prompt P2 does not specify the programming language, and the corresponding answer from ChatGPT is shown in Figure~\ref{fig:prompt_answer_Q2}(d). It is observed that although the prompt P2 does not specify the programming language, the answer to P2 explicitly describes ``This is a Python function'' in the first line. It means that ChatGPT recognized that $c_1$ is written in Python. In addition, although there are some differences overall, the core functional summary in the answer to P2 (i.e., ``This is a Python function that takes a list of integers as input and returns the most frequent even integer in the list.'') is semantically equivalent to that in the answer to P1 (i.e., ``This function takes a list of integers as input and returns the most frequently occurring even number in the list.''). It indicates that ChatGPT can recognize the programming language of the given code itself, so we do not specify the programming language in the following prompts. Based on the above observations, we can conclude that it is feasible to use ChatGPT for code summarization without specifying the programming language.

\subsection{Q3: How to use ChatGPT to generate concise comments?}

From the answers to Q2, we observed a phenomenon that the comments generated by ChatGPT are too long, much longer than the ground-truth comment. Usually, we would like to get comments within one sentence. Usually, human developers do not write comments that long either. Therefore, in this question, we explore how to modify the prompts to guide ChatGPT to generate more concise comments. 

\begin{table*}[htbp]
    \caption{Prompts for Q3 and the corresponding answers by ChatGPT. \#Words: number of words.}
    \label{tab:prompt_answer_Q3}  
    \centering  
    \begin{tabular}{|m{0.4cm}|m{6cm}<{\raggedright}|m{8.5cm}<{\raggedright}|m{1cm}|}
        \hline
        
        No. & Prompt & Answer & \#Words\\
        
        \hline
        
        (1) & Please generate a \bfred{short} comment for the following function:\newline\hspace*{2.5cm}<code>
        & This function finds the most frequent even number in a given list of integers using Python's built-in Counter function. & 19 \\
        
        \hline
        
        (2) & Please generate a \bfred{succinct} comment for the following function:\newline\hspace*{2.5cm}<code>
        & This function finds the most frequent even number in a given list of integers. It utilizes the Counter function and iterates over the items to find the even number with the highest frequency. & 33 \\
        
        \hline
        
        (3) & Please generate a \bfred{concise} comment for the following function:
        \newline\hspace*{2.5cm}<code> & This function returns the most frequent even integer from a given list of integers using Counter and basic logic. & 19 \\
        
        \hline
        
    \end{tabular}
\end{table*}

A naive idea is to add some adjectives with a ``concise'' meaning to the prompt. Specifically, we tried three adjectives, including ``short'', ``succinct'', and ``concise''. Rows (1)--(3) of Table~\ref{tab:prompt_answer_Q3} show the detailed prompts and the corresponding answers by ChatGPT. It is observed that these three comments are significantly shorter than those in Figure~\ref{fig:prompt_answer_Q2} (b) and (d), but are still longer than the ground truth (containing 11 words) shown in Figure~\ref{fig:example_of_code_ground-truth_comment}. Compared with the comment generated using the prompt containing ``concise'' (row (3)), the comment generated using the prompt containing ``short'' (row (1)) is closer to the ground truth.

\subsection{Q4: What kind of prompt does ChatGPT suggest for generating short comments?}

In the previous Q3, We found a seemingly good prompt that can instruct ChatGPT to generate short comments. In this question, we want to what kind of prompt ChatGPT suggests for generating short comments. Table~\ref{tab:prompt_answer_Q4} lists the four prompts we designed for Q4 and the corresponding answers by ChatGPT. 

\begin{table*}[htbp]
    \caption{Prompts for Q4 and the corresponding answers by ChatGPT}
    \label{tab:prompt_answer_Q4}  
    \centering  
    \begin{tabular}{|m{0.5cm}|m{6cm}<{\raggedright}|m{10cm}<{\raggedright}|}
        \hline
        No. & Prompt & Answer \\

        \hline
        
        (1) & If I give you a programming language function and ask you to generate a short comment for it, how would you like me to raise the question? & You could present the programming language function followed by \bfred{a brief explanation of what it does}, and then ask for a concise comment to describe it. For example: ``Here is a JavaScript function that generates a random number between 1 and 10. Can you provide a short comment for it?'' \\
        
        \hline

        (2) & If I give you a programming language function and ask you to generate a short comment for it, how would you like me to prompt? & You can prompt by providing the name of the function and \bfred{a brief description of what it does}. Additionally, any input parameters and their descriptions would also be helpful to generate a more specific comment for the function. \\
        
        \hline

        (3) & If I \bfred{only} give you a function and ask you to generate a short comment for it, how would you like me to raise the question? & You can simply provide the function and ask for a brief description or comment on what the function does or how it works. For example, ``\bfblue{Can you give a short comment on what the reduce() function does in Python?}'' \\

        \hline
        
        (4) & If I ask you to generate a short comment for a programming language function I give you, how would you like me to phrase the question?
        & You could ask me to provide a short comment for a specific programming language function by specifying the programming language and the function name. For example, ``\bfblue{Could you generate a short comment for the Python built-in function `sorted()'?}'' or ``\bfblue{What's a concise description of the JavaScript function `Array.prototype.map()'?}'' \\
        
        \hline
    \end{tabular}
\end{table*}

From rows (1) and (2) of Table~\ref{tab:prompt_answer_Q4}, it is observed that ChatGPT expects the user prompt to provide a brief description of what the input code snippet does. However, this expectation is what humans expect ChatGPT to do. Therefore, the prompts suggested in the answers of rows (1) and (2) are inappropriate. From rows (3) and (4) of Table~\ref{tab:prompt_answer_Q4}, it is observed that ChatGPT provides three examples of suggested prompts (highlighted in blue font).

\subsection{Q5: Which one performs better, the ChatGPT-suggested prompts in Q4 or the prompt proposed in Q3?}

Based on these examples, we design three new prompts. In Table~\ref{tab:prompt_answer_Q5}, row (1) shows the prompt we find in Q3 and the corresponding answers by ChatGPT, while rows (2)--(4) show the three prompts we designed based on the ChatGPT-suggested prompts in Q4.  It is observed that the comments generated using prompts in rows (2)-(4) are significantly longer than that generated using the prompt in row (1).

\begin{table*}[htbp]
    \caption{Prompts and answers by ChatGPT for Q5}
    \label{tab:prompt_answer_Q5} 
    \centering  
    \begin{tabular}{|m{0.5cm}|m{6cm}<{\raggedright}|m{10cm}<{\raggedright}|}
        \hline
        No. & Prompt & Answer \\
        
        \hline
        
         (1) & Please generate a short comment for the following function: \newline\hspace*{2.5cm}<code> & This function finds the most frequent even number in a given list of integers using Python's built-in Counter function. \\
        
        \hline
        \hline

        (2) & Can you give a short comment on what the mostFrequentEven() function does in Python?\newline\hspace*{2.5cm}<code>
        & The mostFrequentEven() function takes a list of integers nums as input and returns the most frequent even number in the list. It uses the Counter function from the Python collections module to count the occurrences of each element in the list. The function then iterates through the counts and checks if the key (i.e., the number) is even and if its count is greater than the current maximum count or if the count is equal to the maximum count and the key is smaller than the current answer. If so, the function updates the answer and maximum count. Finally, the function returns the answer, which is the most frequent even number in the list. If there are no even numbers in the list, the function returns -1. \\

        \hline

        (3) & Could you generate a short comment for the Python function `mostFrequentEven()' \newline\hspace*{2.5cm}<code>
        & The mostFrequentEven() function in Python takes a list of integers as input and returns the most frequent even number in the list. It uses the Counter() function to count the occurrences of each element in the list and iterates through the counts to find the most frequent even number. If there are no even numbers in the list, the function returns -1. \\

        \hline

        (4) & What's a concise description of the Python function `mostFrequentEven()'? \newline\hspace*{2.5cm}<code>
        & mostFrequentEven() is a Python function that takes a list of integers as input and returns the most frequent even number in the list. It uses the Counter() function to count the occurrences of each element in the list and iterates through the counts to find the most frequent even number. If there are no even numbers in the list, the function returns -1. \\
        
        \hline
    \end{tabular}
\end{table*}

\begin{figure}[!t]
  \centering
  \includegraphics[width=0.7\linewidth]{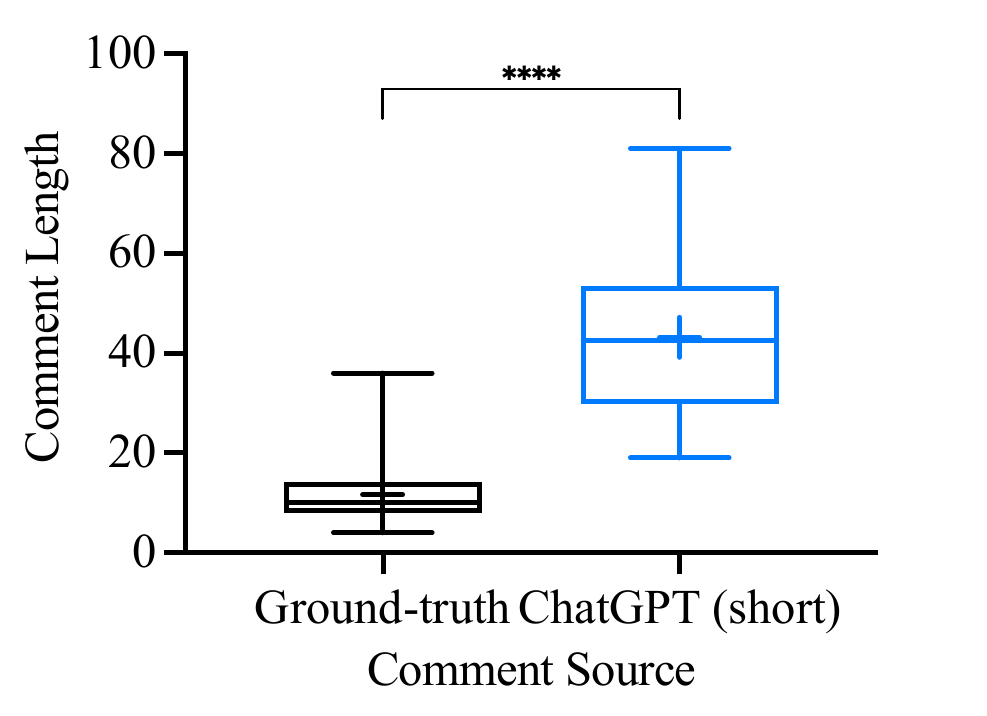}
  \caption{Length of comments generated by ChatGPT and the ground-truth comments. `****' ($p<0,0001$) represents the differences between the two groups are extremely significant.}
  \label{fig:ChatGPT_short_vs_Ground_truth}
\end{figure}

Further, we use more data to verify the prompt containing ``short''. Specifically, we randomly select 50 samples from the CSN-Python test set, then use the prompt containing ``short'' to ask ChatGPT to generate comments. Figure~\ref{fig:ChatGPT_short_vs_Ground_truth} shows the length distribution of the comments generated by ChatGPT using the prompt containing ``short'' (i.e., ChatGPT (short)) and the ground-truth comments (i.e., Ground-truth) on all 50 samples. It is observed that the median, first quartile, and third quartile associated with ChatGPT (short) are larger than those associated with Ground-truth. To test whether there is a statistically significant difference between the two methods, we perform the paired Wilcoxon-Mann-Whitney signed-rank test at a significance level of 5\%, following previously reported guidelines for inferential statistical analysis involving randomized algorithms~\cite{2014-Hitchhiker-Guide-Statistical-Tests}. The $p$-value between ChatGPT (short) and Ground-truth is smaller than the significant threshold value of 0.05, which means the comments generated by ChatGPT (short) are out-of-distribution.

\begin{table*}[htbp]
    \caption{Prompts and the corresponding answers by ChatGPT. \#Words: number of words.}
    \label{tab:prompt_answer_one-sentence_40_words} 
    \centering  
    \begin{tabular}{|m{0.4cm}|m{6cm}<{\raggedright}|m{8.5cm}<{\raggedright}|m{1cm}|}
        \hline
        
        No. & Prompt & Answer & \#Words\\
        
        \hline
        
        (1) & Please generate a short comment \bfred{in one sentence} for the following function:\newline\hspace*{2.5cm}<code> & This function returns the most frequent even number in a given list of integers. & 14 \\
        
        \hline
        \hline
       
        (2) & Please generate a short comment \bfred{within 40 words} for the following function:\newline\hspace*{2.5cm}<code>
        & This function finds and returns the most frequent even integer from a given list of integers. It accomplishes this by utilizing the Counter library and iterating through the list to compare the frequency of even integers. The function returns -1 if there are no even integers in the given list. & 50 \\
        
        \hline
        
        (3) & Please generate a short comment \bfred{in 40 words} for the following function:\newline\hspace*{2.5cm}<code>
        & This function takes in a list of integers and returns the most frequent even integer. It achieves this by using a counter and iterating through the items while checking for even numbers and comparing their frequency. The function is concise and efficient. & 42 \\
        
        \hline
        
        (4) & Please generate a short comment of \bfred{no more than 40 words} for the following function:\newline\hspace*{2.5cm}<code>
        & This Python function takes a list of integers as input and returns the most frequent even number in the list. It uses the Counter module to count the frequency of each number and compares their counts to determine the most frequent even number. & 43 \\
        
        \hline
    \end{tabular}
\end{table*}

We design two new prompts by adding additional constraints on the prompt containing the ``short''. Specifically, we use two types of constraints: sentence limitation and word limitation. We first analyze the data distribution of the ground-truth comments in the CSN-Python test set. 

\begin{figure}[!t]
  \centering
  \includegraphics[width=0.8\linewidth]{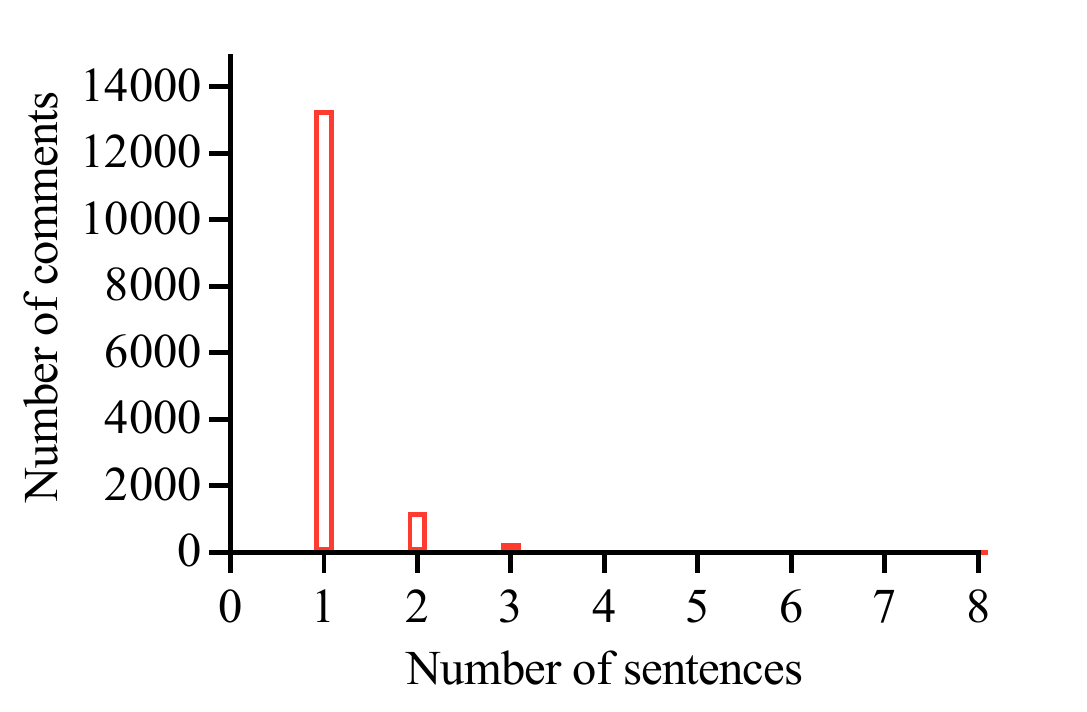}
  \caption{Statistics of the number of sentences in comments}
  \label{fig:comment_number_by_sentence_number}
\end{figure}

Figure~\ref{fig:comment_number_by_sentence_number} shows the statistics of the number of comments according to the number of sentences. It is observed that most ground-truth comments (about 90\%) only contain one sentence.

\begin{figure}[!t]
  \centering
  \includegraphics[width=0.8\linewidth]{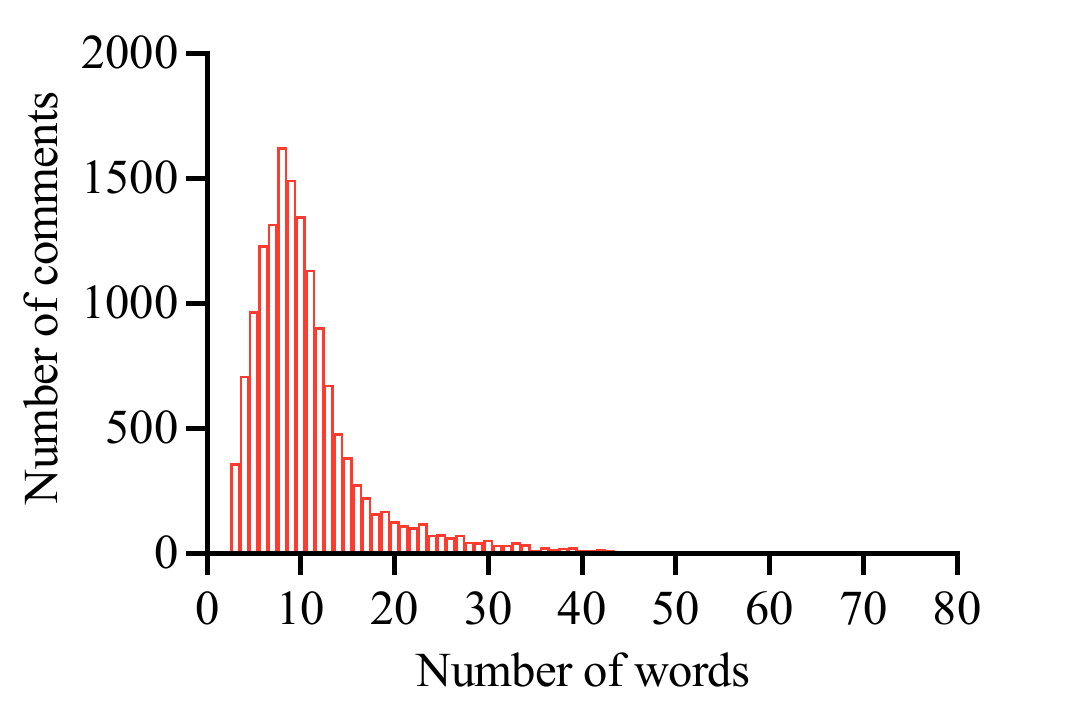}
  \caption{Statistics of the number of words in comments}
  \label{fig:comment_number_by_word_number}
\end{figure}

Figure~\ref{fig:comment_number_by_word_number} shows the statistics of the number of comments according to the number of words they contained. It is observed that most ground-truth comments (about 99\%) consist of less than 40 words.

Row (1) of Table~\ref{tab:prompt_answer_one-sentence_40_words} shows the prompt equipped with the sentence limitation (i.e., ``in one sentence'' highlighted in red font) and the corresponding comment generated by ChatGPT. Rows (2)--(4) show the prompts equipped with the word limitation (i.e., ``within/in/no more than 40 words'' highlighted in red font) and the corresponding comments generated by ChatGPT. Although we explicitly require ChatGPT to generate comments of less than 40 words, it did not understand it.

\begin{figure}[!t]
  \centering
  \includegraphics[width=\linewidth]{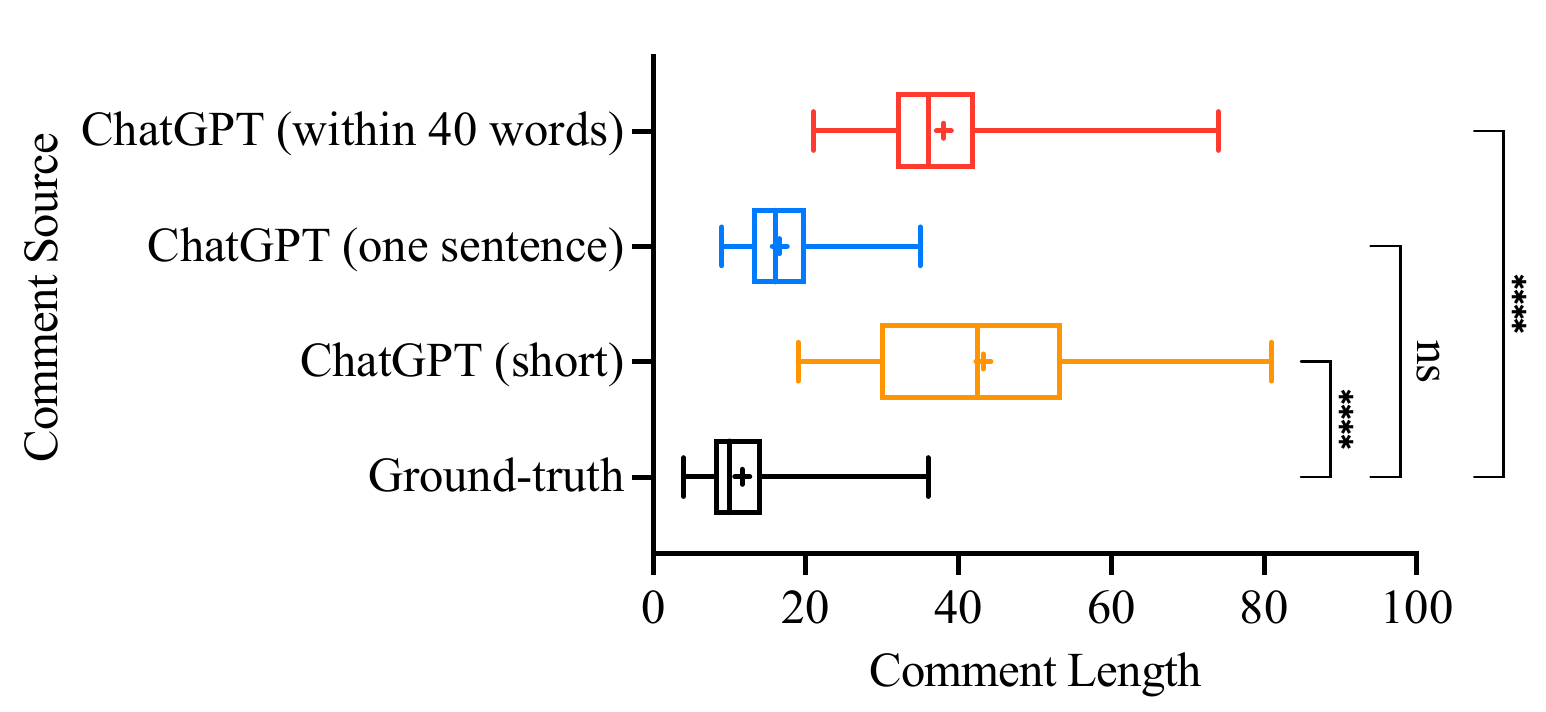}
  \caption{Length distribution of comments on 50 samples. ``*'' ($0.01 < p < 0.05$), ``**" ($0.001 < p < 0.01$), ``***'' ($0.0001 < p < 0.001$) and ``****'' ($p < 0.0001$) represent the differences between two groups are Significant, Very significant, Extremely significant and Extremely significant, respectively. And ‘ns’ ($p \geq 0.05$) means Not significant.}
  \label{fig:compare_length_on_50_sample}
\end{figure}

Further, we use the randomly selected 50 mentioned earlier to verify the prompts containing ``in one sentence'' and ``within 40 words''. Figure~\ref{fig:compare_length_on_50_sample} shows the length distribution of comments generated by ChatGPT using three prompts, where ChatGPT (within 40 words) and ChatGPT (in one sentence) denote that we use the prompts shown in rows (1)--(2) of Table~\ref{tab:prompt_answer_one-sentence_40_words} to guide ChatGPT to generate comments, respectively. From Figure~\ref{fig:compare_length_on_50_sample}, it is observed that except for the comments generated using ChatGPT (one sentence), the length of the comments generated by ChatGPT (within 40 words) and ChatGPT (short) are significantly different from that of the ground-truth comments.

\begin{table}[!t]
    \caption{Performance of ChatGPT with different prompts}
    \label{tab:performance_on_50_samples}
    \begin{tabular}{|l|ccc|}
        \hline
        \multirow{2}{*}{Method} &
        \multicolumn{3}{c|}{CSN-Python} \\
        \cline{2-4}
        
        & BLEU & METEOR & ROUGE-L \\
        
        \hline
        
        ChatGPT (short) & 4.85 & 12.90 & 14.41 \\
        
        ChatGPT (one sentence) & 10.66 & 13.36 & 20.37 \\

        ChatGPT (within 40 words) & 4.83 & 13.01 & 14.05 \\
        
        \hline
    \end{tabular}
\end{table}

We report and compare the quality of the comments generated by ChatGPT (short), ChatGPT (one sentence), and ChatGPT (within 40 words) on all 50 samples. Table~\ref{tab:performance_on_50_samples} shows the performance of ChatGPT with the above three prompts. From this table, it is observed that ChatGPT (one sentence) achieves the highest scores of 10.66 in BLEU, 13.36 in METEOR, and 20.37 in ROUGE-L, and outperforms ChatGPT (short) and ChatGPT (within 40 words).

\begin{center}
    \setlength{\fboxrule}{1pt}
    \fbox{%
      \parbox{0.46\textwidth}{%
        \textbf{Summary.} \\ 
        Based on the carefully crafted heuristic questions and ChatGPT's feedback (including the preliminary results on the randomly selected 50 samples), we find an appropriate prompt that can be used to guide ChatGPT to generate in-distribution comments. The appropriate prompt is ``Please generate a short comment in one sentence for the following function:<code>''.
        }%
    }
\end{center}

\section{Experimental Design}
\label{sec:experimental_design}

\subsection{Dataset}
\label{subsec:dataset}
We conduct experiments on a Python dataset provided by the CodeSearchNet (CSN) corpus~\cite{2019-CodeSearchNet-Challenge}. Lu et al.~\cite{2021-CodeXGLUE} showed that some comments in the CSN corpus contain content unrelated to the code snippets and performed data cleaning on the CSN corpus. Therefore, in this paper, we follow~\cite{2020-CodeBERT, 2021-CodeT5} and use the clean version of the CSN corpus provided by Lu et al.~\cite{2021-CodeXGLUE}, where the training/validation/test set contains 251,820/13,914/14,918 samples, respectively.

\subsubsection{Evaluation Metrics}
\label{subsubsec:evaluation_metrics}
We use three metrics BLEU~\cite{2002-BLEU}, METEOR~\cite{2005-METEOR}, and ROUGE-L~\cite{2004-ROUGE}, to evaluate the quality of comments generated by different models, which are widely used in code summarization~\cite{2022-SCRIPT, 2021-SiT, 2018-Improving-Code-Summarization-via-DRL, 2018-TL-CodeSum, 2017-Transformer, 2016-CODE-NN}. 

\textbf{BLEU}, the abbreviation for BiLingual Evaluation Understudy~\cite{2002-BLEU}, is widely used for evaluating the quality of generated code summaries~\cite{2018-Improving-Code-Summarization-via-DRL, 2018-TL-CodeSum, 2016-CODE-NN}. It is a variant of precision metric, which calculates the similarity by computing the n-gram precision of a generated summary to the reference summary, with a penalty for the overly short length~\cite{2002-BLEU}. It is computed as:
\begin{equation}
	BLEU = BP * exp(\sum_{n=1}^N{w_nlogp_n})
	\label{equ:bleu}
\end{equation}

\begin{equation}
	BP = 
	\begin{cases}
	1, &\text{if $|g| > |r|$} \\
	e^{(1 - \frac{|r|}{|g|})}, &\text{if $|g| \leq |r|$}
	\end{cases}
	\label{equ:bp}
\end{equation}
where $N = 1, 2, 3, 4$ and $w_n = \frac{1}{N}$. $p_n$ is the n-gram precision~\cite{2022-Evaluation-Neural-Code-Summarization}. $BP$ represents the brevity penalty. $g$ and $r$ denote a generated (predicted) summary and a reference summary, respectively. $|g|$ and $|r|$ denote the lengths of $g$ and $r$, respectively. In this paper, we follow~\cite{2021-SiT, 2022-SCRIPT} and show the standard BLEU score which provides a cumulative score of 1-, 2-, 3-, and 4-grams~\cite{2021-Why-My-Code-Summarization-Not-Work}.

\textbf{METEOR}, the abbreviation for Metric for Evaluation of Translation with Explicit ORdering~\cite{2005-METEOR}, is also widely used to evaluate the quality of generated code summaries~\cite{2021-Code-Summarization-for-Smart-Contracts, 2020-Rencos, 2020-RL-Guided-Code-Summarization}. For a pair of summaries, METEOR creates a word alignment between them and calculates the similarity scores. Suppose $m$ is the number of mapped unigrams between the reference summary $r$ and the generated summary $g$, respectively. Then, precision ($P_{unig}$), recall ($P_{unig}$), and METEOR are computed as:

\begin{equation}
	P_{unig} = \frac{m}{|g|},\;\; R_{unig} = \frac{m}{|r|}
	\label{equ:P_unigram}
\end{equation}

\begin{equation}
	METEOR = (1 - \gamma * frag^\beta) * \frac{P_{unig} * R_{unig}}{\alpha * P_{unig} + (1 - \alpha) * R_{unig}}
	\label{equ:meteor}
\end{equation}
where $frag$ is the fragmentation fraction. As in~\cite{2020-Rencos}, $\alpha$, $\beta$, and $\gamma$ are three penalty parameters whose default values are 0.9, 3.0 and 0.5, respectively.

\textbf{ROUGE-L.} ROUGE is the abbreviation for Recall-oriented Understudy for Gisting Evaluation~\cite{2004-ROUGE}. ROUGE-L, a variant of ROUGE, is computed based on the longest common subsequence (LCS). ROUGE-L is also widely used to evaluate the quality of generated code summaries~\cite{2021-Project-Level-Encoding-Code-Summarization, 2021-BASTS, 2021-API2Com}. Specifically, the LCS-based F-measure ($F_{lcs}$) is called ROUGE-L~\cite{2004-ROUGE}, and $F_{lcs}$ is computed as:
\begin{equation}
	R_{lcs} = \frac{LCS(r,g)}{|r|}, \;\; P_{lcs} = \frac{LCS(r,g)}{|g|}
	\label{equ:p_r_lcs}
\end{equation}

\begin{equation}
	F_{lcs} = \frac{(1+\beta^2)R_{lcs}P_{lcs}}{R_{lcs}+\beta^2P_{lcs}}
	\label{equ_f-lcs}
\end{equation}
where $r$ and $g$ also denote the reference summary and the generated summary, respectively. Notice that ROUGE-L is 1 when $g = r$; while ROUGE-L is 0 when $LCS(r,g) = 0$, i.e., which means $r$ and $g$ are completely different. $\beta$ is set to 1.2 as in~\cite{2021-CoCoSum, 2018-Improving-Code-Summarization-via-DRL, 2020-Rencos}.

The scores of BLEU, METEOR, and ROUGE-L are in the range of [0,1] and usually reported in percentages. The higher the scores, the closer the generated summary is to the reference summary, and the better the code summarization performance. All scores are computed by the same implementation provided by~\cite{2020-Rencos}.

\subsection{Baseline}
\label{subsec:baseline}
In this section, we introduce several DL-based code summarization models selected as baselines, including NCS~\cite{2020-Transformer-based-Approach-for-Code-Summarization}, CodeBERT~\cite{2020-CodeBERT}, and CodeT5~\cite{2021-CodeT5}.  

\textbf{NCS.} NCS~\cite{2020-Transformer-based-Approach-for-Code-Summarization} is the first Transformer-based code summarization model. It adopts a Transformer-based encoder-decoder architecture. It incorporates the copying mechanism~\cite{2017-Get-To-Point} in the Transformer to allow both generating words from vocabulary and copying from the source code.

\textbf{CodeBERT.} CodeBERT~\cite{2020-CodeBERT} is a representative pre-trained model for source code. It uses the same model architecture as RoBERTa-base~\cite{2019-RoBERTa}. CodeBERT is trained with the Masked Language Modeling (MLM) task and the Replaced Token Detection (RTD) task. The authors of CodeBERT fine-tune and test it on the code summarization task (also called the code documentation generation task in their paper).

\textbf{CodeT5.} CodeT5~\cite{2021-CodeT5} is the {\sota} pre-trained model for source code. CodeT5 builds on an encoder-decoder framework with the same architecture as T5~\cite{2020-T5}. It is trained with four pre-training tasks, including Masked Span Prediction (MSP) task, Identifier Tagging (IT), Masked Identifier Prediction (MIP), and Bimodal Dual Generation (BDG). Different from CodeBERT, CodeT5 has a pre-trained decoder. The authors of CodeT5 also conduct experiments on the code summarization task.

\textbf{ChatGPT.} Currently, ChatGPT~\cite{2022-ChatGPT} is the most advanced and possibly the most powerful Large Language Model (LLM). 
It is trained with massive texts and codes, including a wide range of user queries, and responses that are coherently relevant to these queries. These training data make ChatGPT applicable to chatbots or virtual assistants. Furthermore, it also employs advanced supervised instruction fine-tuning techniques and RLHF~\cite{2017-Reinforcement-Learning-Human-Preferences} to adapt more effectively to specific tasks or domains. In this paper, we conducted experiments to investigate ChatGPT’s capability on code summarization tasks.

\section{Result Analysis and Case Study}
\label{sec:results}

\subsection{Result Analysis}

\begin{table}[htbp]
    \caption{Performance of ChatGPT and three {\sota} models}
    \label{tab:performance_of_ChatGPT_and_other_NCSum_models}
    \begin{tabular}{|l|ccc|}
        \hline
        \multirow{2}{*}{Method} &
        \multicolumn{3}{c|}{CSN-Python} \\
        \cline{2-4}
        
        & BLEU & METEOR & ROUGE-L \\
        
        \hline
        
        NCS & 15.8 & 10.6 & 31.3 \\
        
        CodeBERT & 18.7 & 12.4 & 34.8 \\

        CodeT5 & 20.0 & 14.7 & 37.7 \\
        
        ChatGPT (one sentence) &  10.28 & 14.40 & 20.81 \\
        
        \hline
    \end{tabular}
\end{table}

In this section, we report and compare the comments generated by ChatGPT (one sentence) and three {\sota} models (including NCS, CodeBERT, and CodeT5) on the entire test set of the CSN-Python dataset. ChatGPT (one sentence) denotes that we uniformly use the following prompt to guide ChatGPT to generate comments for all code snippets in the CSN-Python test set:
\begin{center}
    \small
    Please generate a short comment in one sentence for the following function:
    
    <code>
\end{center}

Table~\ref{tab:performance_of_ChatGPT_and_other_NCSum_models} shows the overall performance of ChatGPT (one sentence), NCS, CodeBERT, and CodeT5. From this table, it is observed that 1) in terms of BLEU and ROUGE-L, ChatGPT obtains the lowest scores of 10.28 in BLEU and 20.81 in ROUGE-L, worse than all three {\sota} baselines; in terms of METEOR, ChatGPT is comparable to CodeT5 and is better than NCS and CodeBERT. 

\begin{figure}[htbp]
  \centering
  \includegraphics[width=\linewidth]{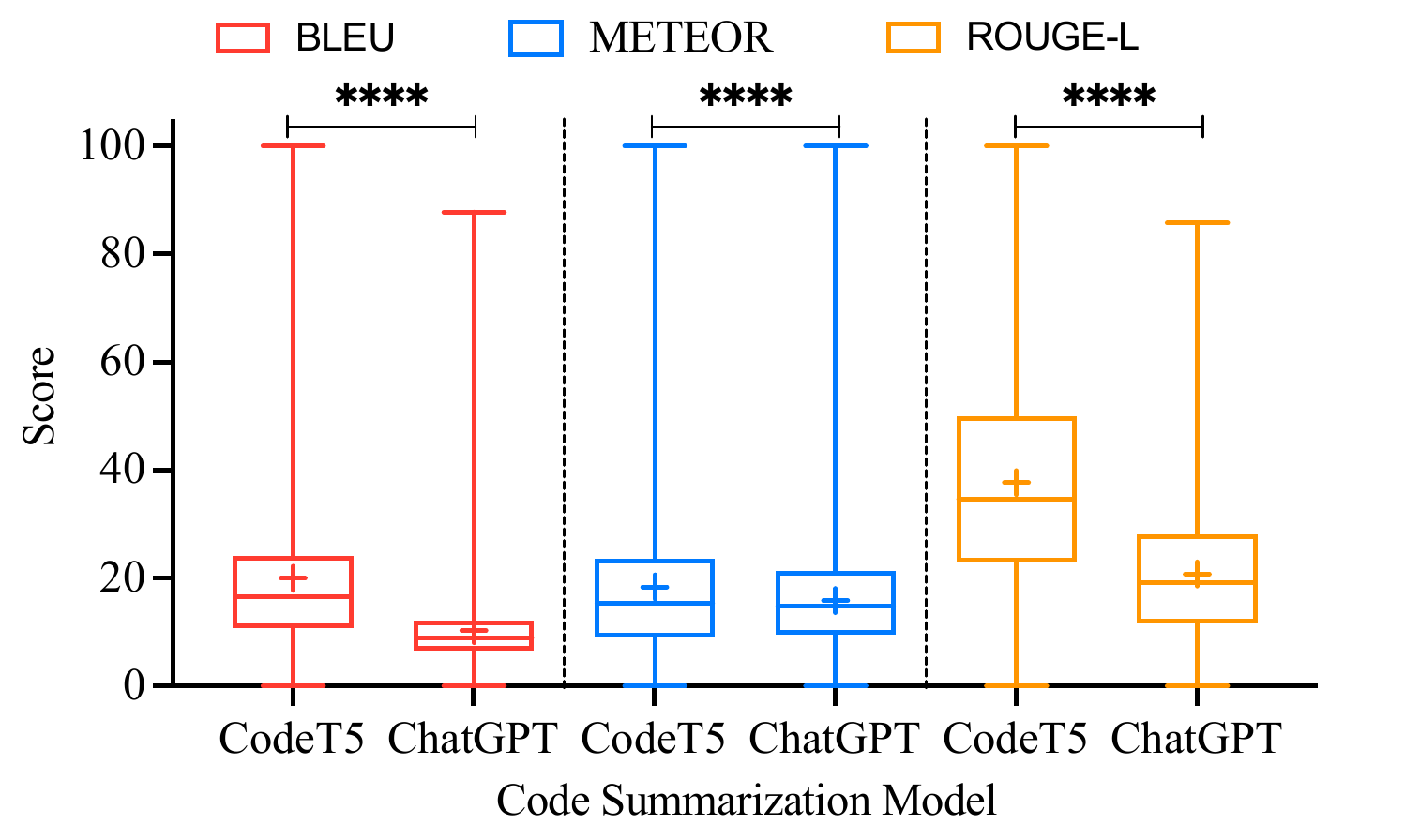}
  \caption{Distribution of metric scored obtained by CodeT5 and ChatGPT (one sentence)}
  \label{fig:compare_ChatGPT_vs_CodeT5(one_sentence)}
\end{figure}

Figure~\ref{fig:compare_ChatGPT_vs_CodeT5(one_sentence)} shows the score distribution of the comments generated by ChatGPT and the best baseline CodeT5 on all samples in the test set of the CSN-Python dataset. It is observed that 1) in terms of BLEU and ROUGE-L, the median, first quartile, and third quartile associated with CodeT5 are larger than those associated with ChatGPT; 2) in terms of METEOR, the third quartile associated with CodeT5 is larger than those associated with ChatGPT. To test whether there is a statistically significant difference between CodeT5 and ChatGPT, we perform the paired Wilcoxon-Mann-Whitney signed-rank test at a significance level of 5\%. All $p$-values between CodeT5  and ChatGPT for all three metrics are smaller than the significant threshold value of 0.05, which means the comments generated by CodeT5  are significantly better than that generated by ChatGPT.

\subsection{Case Study}
In this section, we provide case studies to understand the generated summaries of ChatGPT compared with CodeT5 to demonstrate the strengths and weaknesses of ChatGPT

\subsubsection{successful cases}

\begin{figure}[htbp]
  \centering
  \includegraphics[width=\linewidth]{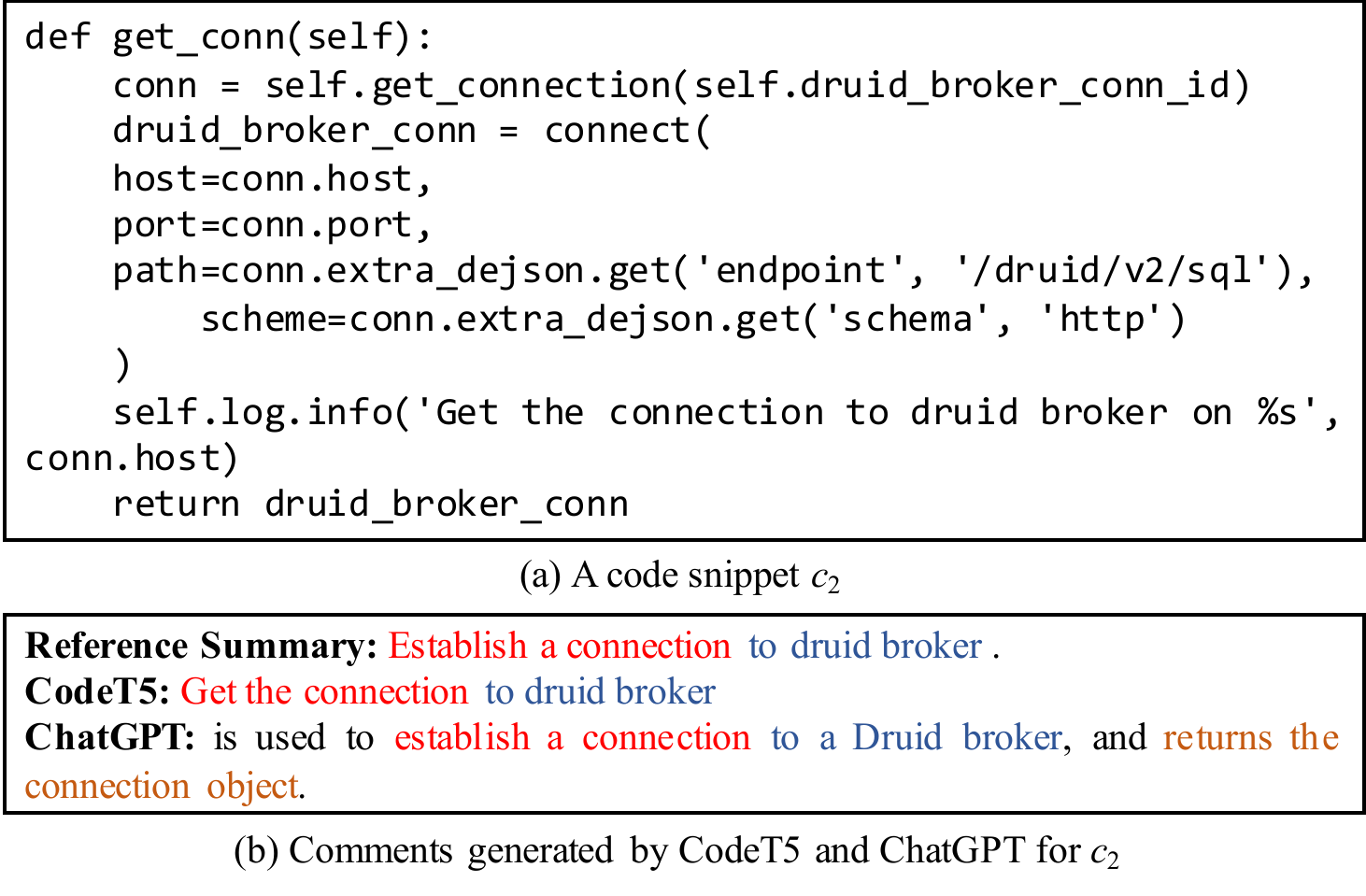}
  \caption{A successful example of ChatGPT}
  \label{fig:successful-case}
\end{figure}

Figure \ref{fig:successful-case}(a) is a code snippet in CSN-Python and Figure \ref{fig:successful-case}(b) shows the reference summary and the comments generated by CodeT5 and ChatGPT. The comment generated by ChatGPT cover the main semantics of ``Establish a connection'' and ``to druid broker'' in ground-truth. But the comment generated by CodeT5 use the wrong verb ``Get'' instead of ``Establish'', which may be caused by the frequent appearance of ``get connection'' in the code. It indicats that ChatGPT can better understand the inner meaning of the code without being interfered by the vocabulary of the code. Besides, ChatGPT tends to describe more detailed behavior of the code in the comments generated, such as ``returns the connection object''.

\subsubsection{moderate cases}

\begin{figure}[htbp]
  \centering
  \includegraphics[width=\linewidth]{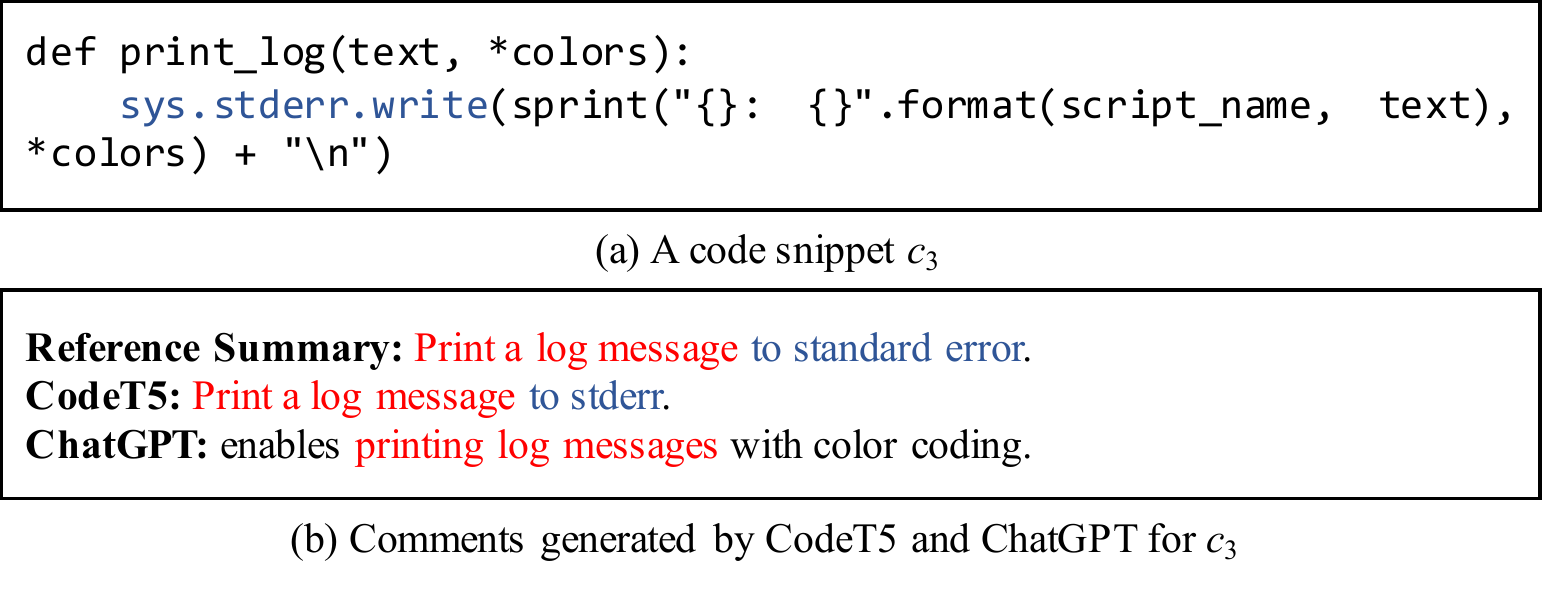}
  \caption{A moderate example of ChatGPT}
  \label{fig:moderate-case}
\end{figure}

Figure \ref{fig:moderate-case}(b) shows the reference summary and the comments generated by CodeT5 and ChatGPT for the code snippet in Figure \ref{fig:moderate-case}(a) (a code snippet in CSN-Python).
The comments generated by CodeT5 and ChatGPT both cover the main semantics of ``Print a log message'' in the reference summary. However, the second part ``to standard error'' is ignored in the comments generated by ChatGPT and is replaced irrelevant information ``with color coding''. It is worth mentioning that the comment generated by CodeT5 contains ``to stderr'', which might have taken a hint from ``sys.stderr.write'' in the code. It demonstrates that, compared with CodeT5, ChatGPT is less sensitive to the features or keywords of a programming language.

\subsubsection{weak cases}
\begin{figure}[htbp]
  \centering
  \includegraphics[width=\linewidth]{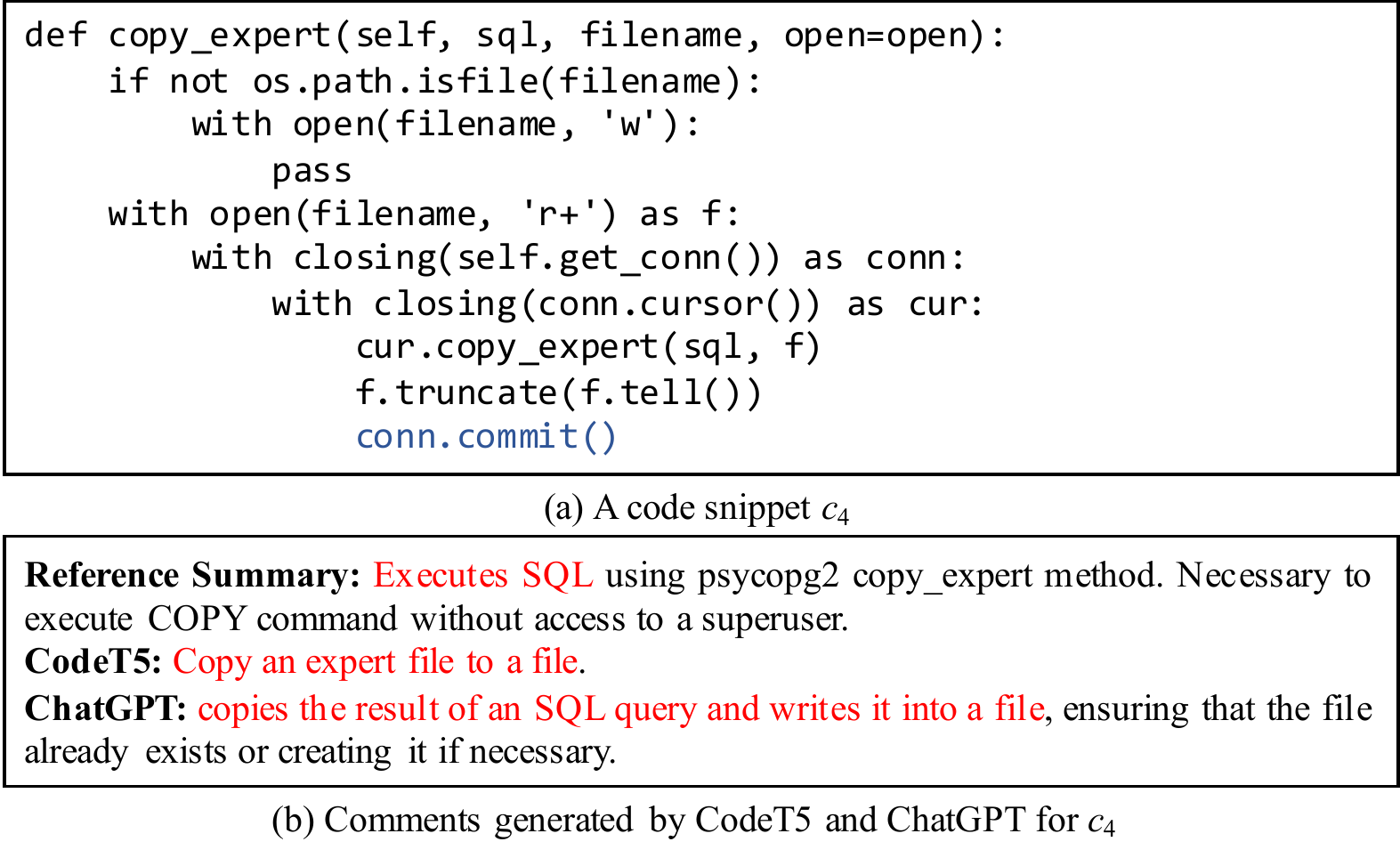}
  \caption{A weak example of ChatGPT}
  \label{fig:weak-case}
\end{figure}

Figure~\ref{fig:weak-case}(a) is a code snippet in CSN-Python. Figure~\ref{fig:weak-case}(b) shows the reference summary and the comments generated by CodeT5 and ChatGPT. Neither CodeT5 nor ChatGPT realized that the function of the code is to ``Execute SQL'' statements. They misunderstood it for the copy operation, which may be caused by ignoring the sentence ``conn.commit()''. It reveals that ChatGPT might still have some difficulty understanding the deep logic of the code.

\section{Related work}
\label{sec:related_work}
The neural machine translation (NMT)-based models are widely used to generate summaries for code snippets with encoder-decoder neural networks~\cite{2016-CODE-NN, 2018-DeepCom, 2019-Ast-attendgru, 2020-Hybrid-DeepCom, 2020-Code-to-Comment-Translation, 2021-BASTS, 2021-SiT, 2022-SCRIPT}. For example, Iyer et al.~\cite{2016-CODE-NN} are the first to apply deep learning to automatic code summarization. They adopt LSTM networks~\cite{1997-LSTM} with attention to leverage the code vectors and generate natural language sentences that describe C\# code snippets and SQL queries. Hu et al.~\cite{2018-TL-CodeSum} use one additional encoder to encode API sequences and improve the summary generation by learned the API knowledge. Subsequently, various additional information is incorporated to further improve DL-based code summarization performance, such as abstract syntax trees~\cite{2018-Improving-Code-Summarization-via-DRL, 2019-Ast-attendgru, 2019-Code-Summarization-with-Extended-Tree-LSTM, 2021-BASTS, 2021-SiT, 2022-SCRIPT}, code property graphs~\cite{2020-FusionGNN}, similar code snippets~\cite{2020-R2Com, 2021-EditSum}, file context~\cite{2020-Improved-Summarization-Attention-File-Context}, etc. Recently, with the success of the pre-training and fine-tuning paradigm in the field of NLP (e.g., BERT~\cite{2019-BERT} and T5~\cite{2020-T5}), many works introduce this paradigm to further boost neural code summarization, such as CodeBERT~\cite{2020-CodeBERT} and CodeT5~\cite{2021-CodeT5}. These works first pre-train a model with general language modeling tasks, such as MLM and ULM. Then, they fine-tune the pre-trained models on code summarization~\cite{2020-CodeBERT, 2021-CodeT5, 2022-UniXcoder}. 

Recently, there are growing interest in leveraging ChatGPT for various software engineering tasks. Sobania et al.~\cite{2023-ChatGPT-For-Automatic-Bug-Fixing} compare the performance of ChatGPT with the performance of Codex and several dedicated automatic program repair (APR) approaches. They claim that human input can be of much help to an automated APR system, but it requires a mental cost to verify ChatGPT answers. They propose that incorporation of automated approaches to provide ChatGPT with hints as well as automated verification of its responses would yield ChatGPT to be a viable tool. Cao et al.~\cite{2023-ChatGPT-for-Program-Repair} study ChatGPT’s ability in DL program debugging. They explore the impact of prompts on the debugging performance and propose a prompt template to improve the performance. Additionally, they explore how far ChatGPT’s dialogue feature can push the performance forward. They summarize the pros and cons of ChatGPT’s reply and point out the possible directions where ChatGPT could help to facilitate the SE community. Tian et al.~\cite{2023-Is-ChatGPT-the-Ultimate-Programming-Assistant} investigate ChatGPT on three code-related tasks: code generation, program repair, and code summarization. They find out that ChatGPT struggles to generalize to new and unseen problems and long prompts have a negative impact on the inference capabilities of ChatGPT. They also discover that ChatGPT has a limited attention span.

\section{Threats to Validity}
\label{sec:threats_to_validity}

\textbf{Threats to External Validity.} ChatGPT generate varied answers for identical code snippets with different prompts due to their inherent randomness. One threat to the validity of our study is that conclusions drawn from random results may be misleading. To mitigate this threat, we conducted a pre-study to determine the prompt so that the comments generated by ChatGPT are more like real-life comments. In addition, we conduct our experiment on a large dataset, which helps reduce the impact of randomness to some extent. Another potential threat to validity is the selection of the programming language and benchmark dataset. The results might vary depending on the programming language and size of the benchmark dataset. To mitigate these threats, we chose a popular large benchmark dataset (i,e., the CSN corpus) and targeted Python -- the most popular programming language. We will explore the code summarization performance of ChatGPT on other programming language datasets provided in the CodeSearchNet corpus in future work.

\textbf{Threats to Construct Validity.} ChatGPT is currently under active development. The ChatGPT we evaluate is based on the GPT-3.5 model. During our study there was a major update to it, the latest model is GPT-4. Therefore, the results in this paper might have been an underestimation of the latest ChatGPT. In future studies, we plan to investigate the GPT-4-based ChatGPT and compare its performance with that of previous versions.

\section{Opportunities}
\label{sec:opportunities}

\begin{figure*}[htbp]
  \centering
  \includegraphics[width=\linewidth]{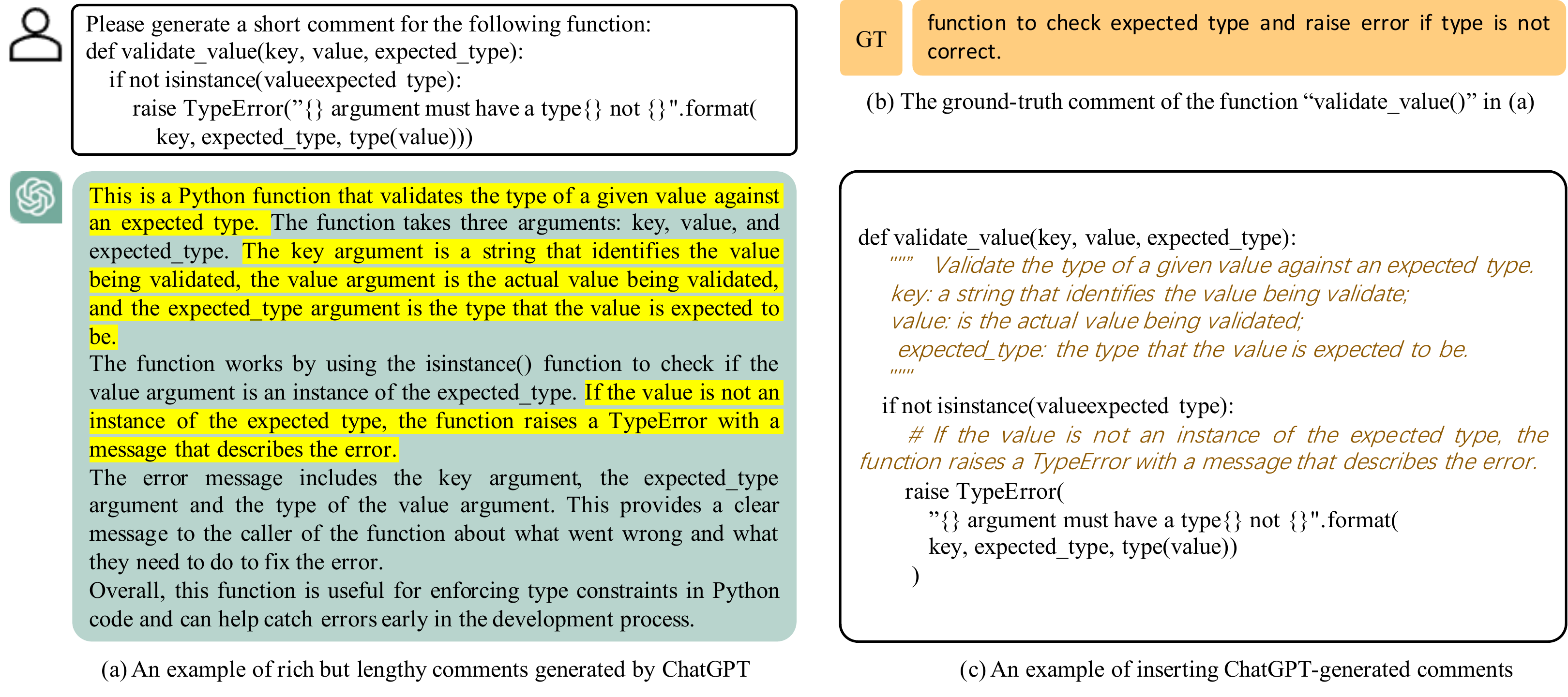}
  \caption{An example of cropping and templating comments generated by ChatGPT}
  \label{fig:example_of_cropping_templating_comment}
\end{figure*}

\noindent\textbf{(1) A lot of room for improvement on existing datasets.} 

In this paper, we directly run ChatGPT on the CSN-Python test set without any fine-tuning process. This is a zero-shot application scenario. From the results in Table~\ref{tab:performance_of_ChatGPT_and_other_NCSum_models}, it is observed that ChatGPT's zero-short code summarization performance still has a lot of room for improvement on the CSN-Python dataset. Since the large language model GPT 3.5/4 behind ChatGPT is not open-sourced, fine-tuning it is not practical. 
A plausible way to improve its code summarization performance is to explore more efficient prompts. However, it is time-consuming and labor-intensive to rely on manual attempts. How to use prompt engineering to automatically explore better prompts is worthy of further study. In this paper, we provide two optimization directions to explore appropriate prompts to guide ChatGPT to generate in-distribution comments, i.e., sentence limitation and word limitation. Other optimization directions are also worthy of further exploration.

\noindent\textbf{(2) Cropping and templating the rich but lengthy comments that ChatGPT generates by default.}

In Section~\ref{sec:ChatGPT_for_code_summarization}, we find that ChatGPT can generate rich and lengthy comments. It should be noted that although the comments generated by ChatGPT are relatively long, it does not mean that these long comments are useless. Figure~\ref{fig:example_of_cropping_templating_comment} shows an example, where (a) shows the prompt and comments generated by ChatGPT for the function ``validate\_value()''; (b) shows the ground-truth comment of the function ``validate\_value()''; and (c) shows the example of inserting the cropping and templating comments generated by ChatGPT into the function ``validate\_value()''. From the figure, it is observed that 1) compared with the ground-truth comment, the comments generated by ChatGPT contain more details; 2) cropping the comments generated by ChatGPT and wrapping the cropped comments (e.g., the comments highlighted in yellow shown in Figure~\ref{fig:example_of_cropping_templating_comment}(a)) with a template (e.g., the docstrings inserted in Figure~\ref{fig:example_of_cropping_templating_comment}(c)) are very useful and friendly for developers. Therefore, how to automatically extract concise and useful information from such comments is also worthy of further study. Python enhancement proposals~\cite{2001-Python-Enhancement-Proposals} provide detailed coding conventions, which can be used to guide the design of the comment template.

\noindent\textbf{(3) Creating a new benchmark dataset.}

\begin{figure}[t]
  \centering
  \includegraphics[width=\linewidth]{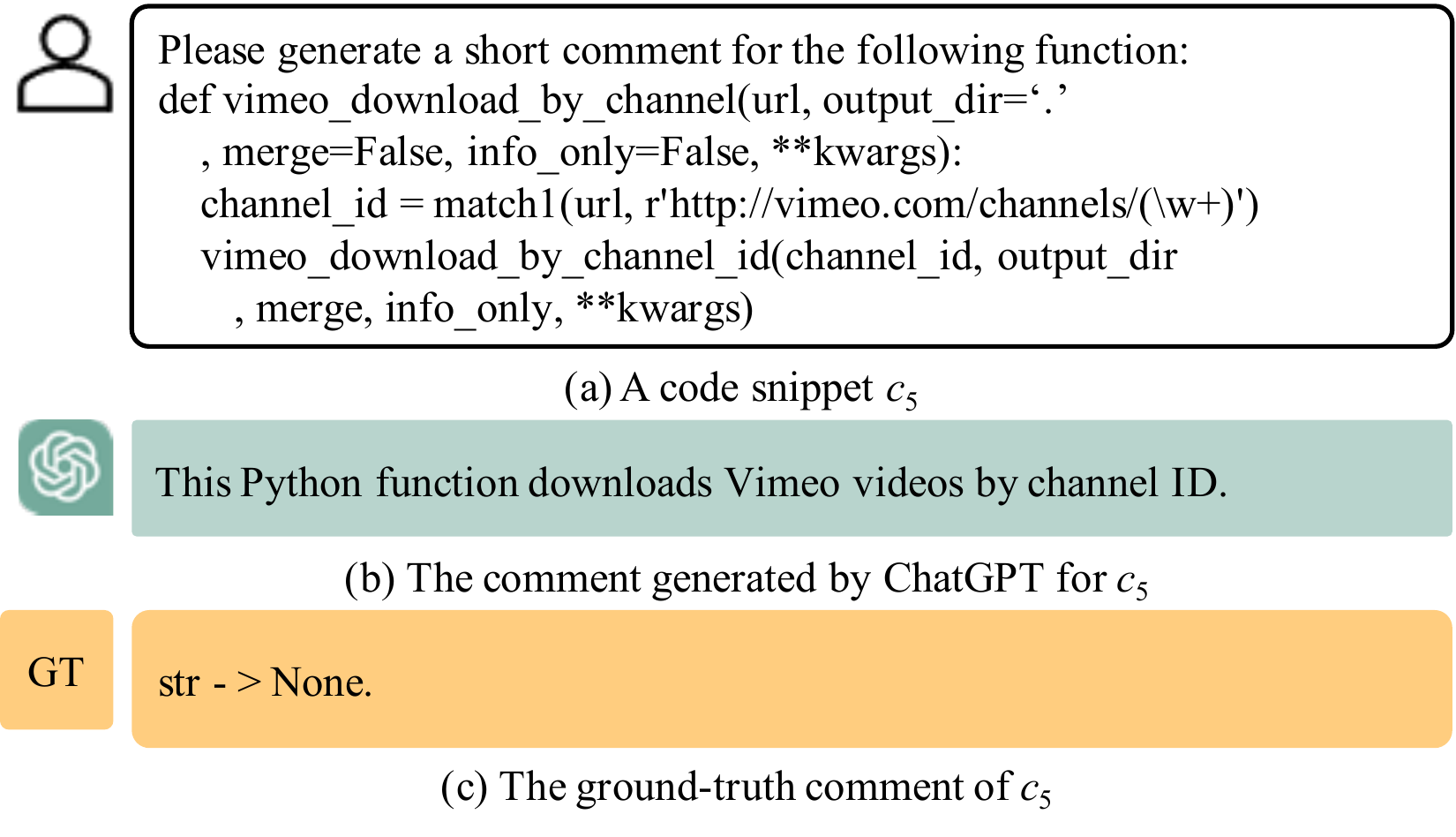}
  \caption{An example of a low-quality ground-truth comment}
  \label{fig:example_of_bad_ground-truth}
\end{figure}

Currently, the quality of datasets (e.g., the CSN corpus~\cite{2019-CodeSearchNet-Challenge}) widely used in code summarization is unverified. These data sets are usually crawled from open-source platforms (such as GitHub~\cite{2008-GitHub}) and then subjected to simple data cleaning. We find some ground-truth comments are low-quality. Figure~\ref{fig:example_of_bad_ground-truth} shows an example of a low-quality ground-truth comment. From Figure~\ref{fig:example_of_bad_ground-truth}(c), it is observed that the ground-truth comment of $c_5$ is too short and does not summarize the functionality of $c_5$. Compared with the ground-truth comment, the comment generated by ChatGPT for $c_5$ is more informative and summarizes the core semantics of $c_5$. Therefore, we believe that how to leverage ChatGPT to construct a high-quality benchmark dataset for code summarization is worth further exploration.

\noindent\textbf{(4) Designing new metrics to evaluate code summarization models.}

\begin{figure}[t]
  \centering
  \includegraphics[width=\linewidth]{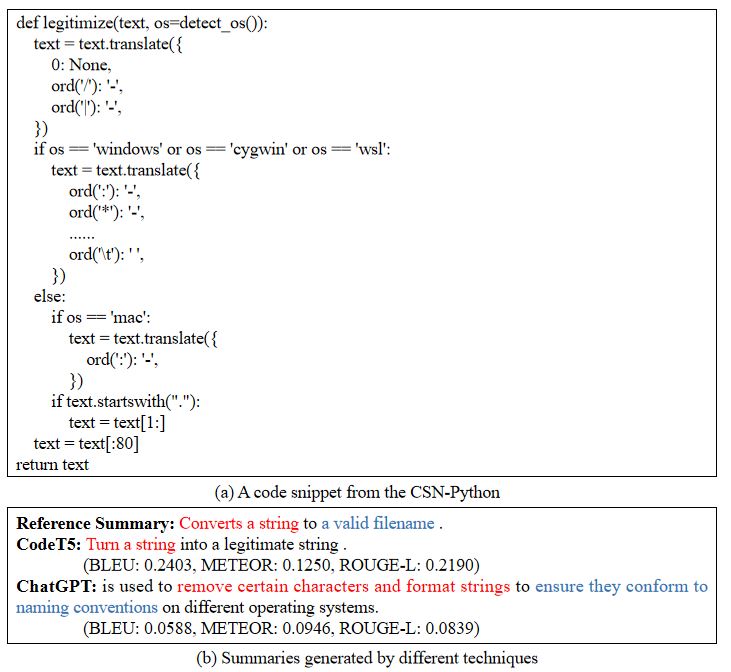}
  \caption{An example of traditional metrics not suitable for evaluating the quality of the comments generated by ChatGPT}
  \label{fig:opportunities-metric}
\end{figure}

Figure~\ref{fig:opportunities-metric}(a) shows a code snippet $c_6$, and Figure~\ref{fig:opportunities-metric}(b) shows the ground-truth comment and the comments generated by CodeT5 and ChatGPT. From the figure, it is observed that although the comment generated by CodeT5 outperforms the one generated by ChatGPT in BLEU, METEOR and ROUGE-L, the semantic ``a valid filename'' is missing in the comment generated by CodeT5. However, the comment generated by ChatGPT covers the semantics of ``Converts a string'' and ``to a valid filename'' though not using the exact words. This is strong evidence that traditional evaluation metrics are no longer suitable for evaluating the quality of the comments generated by ChatGPT.

\section{Conclusion}
\label{sec:conclusion}
In this paper, we preliminarily design heuristic questions to explore an appropriate prompt that can guide ChatGPT to generate in-distribution comments. We compared the automatic code summarization performance of ChatGPT with that of three dedicated code summarization models (i.e., NCS, CodeBERT, and CodeT5) on the CSN-Python dataset. We find that overall ChatGPT is inferior to dedicated code summarization models on the CSN-Python dataset in terms of BLEU, METEOR, and ROUGE-L. Based on our findings, we outline four open challenges and opportunities for applying ChatGPT to code summarization. We hope our results and observations will be helpful for future work with ChatGPT in the field of code summarization.


\bibliographystyle{ACM-Reference-Format}
\bibliography{reference}

\end{document}